\documentclass[letterpaper]{JHEP3}

\usepackage{graphicx}
\usepackage{amsmath}

\title{Nonlinearly charged Lifshitz black holes for any exponent $z>1$}

\author{Abigail Alvarez\\
Departamento~de~F\'{\i}sica,~CINVESTAV--IPN,%
~A.P.~14-740,~07000,~M\'exico~D.F.,~M\'exico.\\
Instituto~de~Ciencias~F\'isicas~y~Matem\'aticas,%
~Universidad~Austral~de~Chile,~Valdivia,~Chile.\\
E-mail: \email{aalvarez-at-fis.cinvestav.mx}}

\author{Eloy Ay\'on--Beato\\
Departamento~de~F\'{\i}sica,~CINVESTAV--IPN,%
~A.P.~14-740,~07000,~M\'exico~D.F.,~M\'exico.\\
Instituto~de~Ciencias~F\'isicas~y~Matem\'aticas,%
~Universidad~Austral~de~Chile,~Valdivia,~Chile.\\
E-mail: \email{ayon-beato-at-fis.cinvestav.mx}}

\author{Hern\'{a}n A. Gonz\'{a}lez\thanks{Laurent Houart postdoctoral fellow.}\\
Physique Th\'eorique et Math\'ematique,
Universit\'e Libre de Bruxelles \& International Solvay Institutes,
Campus Plaine C.P. 231, B-1050 Bruxelles, Belgium.\\
E-mail: \email{hgonzale-at-ulb.ac.be}}

\author{Mokhtar Hassa\"{\i}ne\\
Instituto de Matem\'atica y F\'{\i}sica, Universidad de Talca,
Casilla 747, Talca, Chile.\\
E-mail: \email{hassaine-at-inst-mat.utalca.cl}}

\abstract{Charged Lifshitz black holes for the
Einstein-Proca-Maxwell system with a negative cosmological
constant in arbitrary dimension $D$ are known only if the
dynamical critical exponent is fixed as $z=2(D-2)$. In the
present work, we show that these configurations can be extended
to much more general charged black holes which in addition
exist for any value of the dynamical exponent $z>1$ by
considering a nonlinear electrodynamics instead of the Maxwell
theory. More precisely, we introduce a two-parametric nonlinear
electrodynamics defined in the more general, but less known,
so-called $(\mathcal{H},P)$-formalism and obtain a family of
charged black hole solutions depending on two parameters. We
also remark that the value of the dynamical exponent $z=D-2$
turns out to be critical in the sense that it yields
asymptotically Lifshitz black holes with logarithmic decay
supported by a particular logarithmic electrodynamics. All
these configurations include extremal Lifshitz black holes.
Charged topological Lifshitz black holes are also shown to
emerge by slightly generalizing the proposed electrodynamics.}

\begin{document}

\section{Introduction}

In the last five years, there has been an intensive activity in
order to extend the standard Anti de Sitter/Conformal Field
Theory (AdS/CFT) correspondence \cite{Maldacena:1997re} to the
non-relativistic physics. In this context, the AdS group is
replaced by its non-relativistic cousin the Schr\"odinger group
\cite{Son:2008ye,Balasubramanian:2008dm} or by the Lifshitz
group \cite{Kachru:2008yh}. These two groups share in common an
anisotropic scaling symmetry that allows the time and space
coordinates to rescale with different weight
\begin{equation}\label{eq:anisyymetry}
t\mapsto\lambda^z\,t, \qquad \vec{x}\mapsto\lambda\vec{x},
\end{equation}
where the constant $z$ is called the dynamical critical
exponent. The main difference between the Schr\"odinger and the
Lifshitz groups lies in the fact that the former enjoys in
addition a Galilean boost symmetry whose gravity dual metric
requires the introduction of a closed null direction in
addition to the holographic direction. This extra coordinate is
hard to understand holographically which makes its physical
interpretation a subject of debate. In the present work, within
the spirit of the AdS/CFT correspondence, we are interested in
looking for black hole solutions whose asymptotic metrics enjoy
the anisotropic symmetry (\ref{eq:anisyymetry}) without the
Galilean boost symmetry. In this case, the zero temperature
gravity dual is the so-called Lifshitz spacetime defined by the
metric \cite{Kachru:2008yh}
\begin{equation}\label{eq:Lifshitz}
ds^2=-\frac{r^{2z}}{l^{2z}}dt^2+\frac{l^2}{r^2}dr^2+r^2d\vec{x}^2,
\end{equation}
where $\vec{x}$ is a $(D-2)$-dimensional vector. Lifshitz black
holes refer to black hole configurations whose metrics
reproduce asymptotically the Lifshitz spacetime and their role
holographically is to capture the non-relativistic behavior at
finite temperature. This is the lesson learned from the AdS/CFT
correspondence, which is recovered for $z=1$, and that has been
proved to be a promising tool for studying strongly correlated
quantum systems. However, other values of $z$ are present
experimentally in several problems of condensed matter physics,
and it is then natural to explore the existence of Lifshitz
black holes with the aim of widening the scope of holographic
applications. From the very beginning of the Lifshitz advent it
becomes clear that pure Einstein gravity with eventually a
cosmological constant resists to support such configurations,
since it is clear from Birkhoff's theorem and its
generalizations that in highly symmetric vacuum situations
there is no freedom to choose the boundary conditions and they
become fixed by the dynamics. Consequently, Lifshitz asymptotic
requires the introduction of matter sources
\cite{Taylor:2008tg,Lif-M} or/and to consider higher-curvature
gravity theories \cite{Lif-R2,Matulich:2011ct,Oliva:2012zs}.
One of the first and simplest
examples that has been used for supporting the Lifshitz
spacetimes at any dimension $D$ is a Proca field coupled to
Einstein gravity with a negative cosmological constant
\cite{Taylor:2008tg}. However, as we know from the classical
days of the ``no-hair'' conjecture Proca fields are not enough
to produce black holes by itself due to its impossibility to
make non-trivial contributions to global charges due to its
massive behavior. This is not the case for the massless Maxwell
field and electrically charged black holes are known at any
dimension, however, this gauge field along is incompatible with
the Lifshitz asymptotic. Therefore, the superposition of both
spin-$1$ fields looks a priori as a promising strategy. In
fact, charged black hole solutions which are asymptotically
Lifshitz can be constructed adding the Maxwell action to the
described Proca system \cite{Pang:2009pd}. Unfortunately, up to
now, the only configurations that have been obtained are those
with the single value $z=2(D-2)$ for the dynamical exponent at
any dimension \cite{Pang:2009pd}.

Here, we prove that the charged Lifshitz black holes found for
$z=2(D-2)$ in \cite{Pang:2009pd} can be extended to much more
general black hole configurations which are in addition
characterized by any value of the dynamical exponent $z>1$.
This task is achieved considering a nonlinear version of the
electrodynamics instead of the Maxwell theory. The interest on
nonlinear electrodynamics started with the pioneer work of Born
and Infeld whose main motivation was to cure the problem of
infinite energy of the electron \cite{Born:1934gh}. However,
since the Born-Infeld model was not able to fulfill all the
hope, nonlinear electrodynamics became less popular until
recent years, although outstanding work was done in the middle
\cite{Plebanski:1968}. The renewal interest on nonlinear
electrodynamics matches with their emergence and importance in
the low energy limit of heterotic string theory. Also,
nonlinear electrodynamics have been proved to be a powerful and
useful tool in order to construct black hole solutions with
interesting features and properties as for example regular
black holes \cite{AyonBeato:1998ub} or black holes with
nonstandard asymptotic behaviors in Einstein gravity or some of
its generalizations \cite{Fq}. Charged black hole solutions
emerging from nonlinear theories also have nice thermodynamics
properties which make them attractive to be studied. Usually,
the presence of nonlinearities render the equations much more
difficult to handle, and the hope to obtain exact analytic
solutions interesting from a physical point of view may be
considerably restricted. However, as shown below the derivation
of these charged black hole solutions is strongly inherent to
the presence of the nonlinearities. To be more precise, we
introduce a two-parametric nonlinear electrodynamics defined in
the more general, but less known, so-called
$(\mathcal{H},P)$-formalism and obtain a family of charged
black hole solutions depending on two parameters and valid for
any exponent $z>1$.

The plan of the paper is organized as follows. In the next
section, we present the $(\mathcal{H},P)$-formalism together
with the new proposed electrodynamics, the action including the
Proca system and derive the corresponding field equations. In
Sec.~$3$, we consider the simplest Lifshitz black hole ansatz,
as well as electric ansatze for the spin-$1$ fields and solve
the field equations. We exhibit a family of two-parametric
charged black hole solutions for a generic value of the
dynamical exponent $z>1$ and $z\not=D-2$. We prove that this
last  value of the dynamical exponent $z=D-2$ turns out to be
critical in the sense that it yields asymptotically Lifshitz
black holes with logarithmic decay supported by a particular
logarithmic electrodynamics which can be obtained as a
nontrivial limit of the initially proposed one. We also
construct charged topological Lifshitz black holes by slightly
generalizing the proposed electrodynamics. In Sec.~$4$, we
analyze in detail a particular case of our
$(\mathcal{H},P)$-electrodynamics with invertible constitutive
relations, for which the nonlinear electrodynamics Lagrangian
can be written as a single function of the Maxwell invariant of
the strength $F_{\mu\nu}$, concretely becoming a power law
intensively studied in the recent literature, see
e.g.~\cite{Fq}. In Sec.~$5$, we exhaustively study the
existence of event horizons for all the values of the
parameters. We conclude that there are eight kinds of different
black hole families depending on the structural coupling
constant of the electrodynamics and the dynamical critical
exponent $z$. We present graphs for each of these families in
order to make transparent the involved arguments. Finally, the
last section is devoted to our conclusions and further works.

\section{Action and field equations}

We are interested in extending  the charged Lifshitz black
holes obtained in \cite{Pang:2009pd} for a particular value of
the dynamical exponent to much more general black hole
configurations valid for any value $z>1$. We show that this
task can be achieved by allowing the electrodynamics to behave
nonlinearly. More precisely, in addition to the Proca field
that supports the anisotropic asymptotic in the standard
gravity we incorporate an appropriate nonlinear
electrodynamics, see \cite{Plebanski:1968} for a clever
exposition of the formalism. The dynamics of this theory is
governed by the following action
\begin{eqnarray}
S[g^{\mu\nu},B_\mu,A_\mu,P^{\mu\nu}]&=&\int{d}^Dx\sqrt{-g}\left[\frac1{2\kappa}(R-2\lambda)
-\frac{1}{4}H_{\mu\nu}H^{\mu\nu}-\frac{1}{2}m^2B_{\mu}B^{\mu}\right.\nonumber\\
&&\qquad\qquad\qquad{}\left.-\frac12P^{\mu\nu}F_{\mu\nu}+\mathcal{H}(P)
\right].
\label{eq:action}
\end{eqnarray}
Here, $R$ stands for the scalar curvature of the metric
$g_{\mu\nu}$ and $\lambda$ is the cosmological constant. The
tensor $H_{\mu\nu}=\partial_{\mu}B_{\nu}-\partial_{\nu}B_{\mu}$
is the field strength of the Proca field $B_{\mu}$ having mass
$m$ and whose field equation is given by
\begin{subequations}\label{eq:EPNLE}
\begin{equation}\label{eq:Proca}
\nabla_\mu H^{\mu\nu}=m^2B^\nu.
\end{equation}
The last two terms in the action describe the nonlinear
behavior of the electromagnetic field $A_{\mu}$ with field
strength
$F_{\mu\nu}=\partial_{\mu}A_{\nu}-\partial_{\nu}A_{\mu}$. The
origin of these terms is the following. The naive way to
introduce nonlinear effects in electrodynamics is to consider
a Lagrangian given by a  general function of the quadratic invariants
build out from the strength field $F_{\mu\nu}$. However, a different
formulation is to rewrite this Lagrangian in terms of a
Legendre transform $\mathcal{H}(P)$, function of the invariants
formed with the conjugate antisymmetric tensor $P^{\mu\nu}$, namely
$P\equiv\frac{1}{4}P_{\mu\nu}P^{\mu\nu}$. The
resulting action turns out to be also a functional  of the conjugate
antisymmetric tensor $P^{\mu\nu}$ in addition to the
electromagnetic field $A_{\mu}$, and is given by the last two
terms of (\ref{eq:action}). The advantage of this formulation
lies in the fact that the variation with respect to the electromagnetic field
$A_{\mu}$ gives the nonlinear version of the Maxwell equations
\begin{equation}\label{eq:Max}
\nabla_\mu P^{\mu\nu}=0,
\end{equation}
and these equations are now simply  expressed in terms of the
antisymmetric tensor $P^{\mu\nu}$. These latter can also be
easily integrated and from the expression, the strength field of  the
original electromagnetic field can be calculated using now the
variation of the action with respect to the conjugate
antisymmetric tensor $P^{\mu\nu}$, which gives the so-called
constitutive relations
\begin{equation}\label{eq:consti}
F_{\mu\nu} = \mathcal{H}_P P_{\mu\nu},
\end{equation}
where $\mathcal{H}_P\equiv\partial\mathcal{H}/{\partial P}$ and
the structural function $\mathcal{H}(P)$ defines the concrete
electrodynamics. Note that the standard linear Maxwell theory is
recovered for $\mathcal{H}(P)=P$. Obviously, this formulation
resembles the standard electrodynamics in a media in four
dimensions where the components of the antisymmetric tensors
$P_{\mu\nu}$ and $F_{\mu\nu}$ are just the components of the
vector fields $(\vec{D},\vec{H})$ and $(\vec{E},\vec{B})$,
respectively. In this sense a given nonlinear electrodynamics
is equivalent to a given media, and then from this analogy it is clear
that the components of the tensor $P_{\mu\nu}$ are the appropriate
field strengths to describe nonlinear electrodynamics. It is
important to emphasize that in many cases both formalisms are
not equivalent since as it occurs in standard Legendre transforms, the
equivalence depends on the invertibility of the conjugate
relations, which for electrodynamics are embodied in the
constitutive relations (\ref{eq:consti}). In fact, there are
well behaved nonlinear electrodynamics where the Lagrangian
cannot be written as a single function of the invariants of
$F_{\mu\nu}$; relevant examples are those giving rise to
regular black holes \cite{AyonBeato:1998ub}. This is also the
case we consider here. Hence, the use of the
described formalism is compulsory since the precise nonlinear
electrodynamics we propose as source for the Lifshitz black
holes with generic dynamical critical exponent $z$ is
determined by the following structural function
\begin{eqnarray}
\mathcal{H}(P)&=&-
\frac{[2z^2-Dz+2(D-2)]}{2\kappa l^2}\beta_1\sqrt{-2l^2P}
-\frac{(z-1)(z-D+2)^2}{\kappa z}{\beta_1}^2P \nonumber\\\nonumber\\
&&{}+\frac{(D-2)^2}{2\kappa l^2}\beta_2\left(-2l^2P\right)^{\frac{z}{2(D-2)}},\label{eq:H}
\end{eqnarray}
where $\beta_1$ and $\beta_2$ are dimensionless coupling
constants for any dimensionless election of the Einstein
constant $\kappa$. The invariant $P$ is negative definite since
we will only consider purely electric configurations.

The Einstein equations resulting from varying the action with respect to the metric  yield
\begin{eqnarray}
G_{\mu\nu} + \lambda g_{\mu\nu} &=& \kappa \biggl[
H_{\mu\alpha}H_\nu^{~\alpha}-\frac14g_{\mu\nu}H_{\alpha\beta}H^{\alpha\beta} +
m^2\left(B_\mu B_\nu-\frac12g_{\mu\nu}B^\alpha B_\alpha\right) \nonumber\\
&&{}\quad+\mathcal{H}_P P_{\mu\alpha}P_\nu^{~\alpha}-g_{\mu\nu}(2P\mathcal{H}_P-\mathcal{H}) \biggr].
\label{eq:Ein}
\end{eqnarray}
\end{subequations}
The Einstein-Proca-Nonlinear Electrodynamics system
(\ref{eq:EPNLE}) is the one we pretend to solve in the context
of a Lifshitz asymptotic with generic anisotropy. A last remark
on this subject, usually a nonlinear electrodynamics is
considered plausible if it satisfies the correspondence
principle of approaching the Maxwell theory
($\mathcal{H}\approx{P}$) in the weak field limit
($|l^2P|\ll{1}$), which is obviously not satisfied for our
proposal (\ref{eq:H}). However, this expectation is realized
for self-gravitating charged configurations if the simultaneous
weak field limit of the gravitational background is flat
spacetime, where the Maxwell theory is phenomenologically
viable, or another spacetime where at least there are no
theoretical obstructions for the behavior of standard
electrodynamics. We emphasize that Lifshitz asymptotic with
generic anisotropy is apparently out of this category, since
Lifshitz black holes charged in standard way \cite{Pang:2009pd}
are known only if the dynamical critical exponent is fixed to
$z=2(D-2)$. Notice this is precisely the exponent for which
Maxwell theory is recovered in our electrodynamics (\ref{eq:H})
if additionally the structural constants are chosen as
$\beta_1=0$ and $\beta_2$ is appropriately fixed. Summarizing,
the proposed electrodynamics has no correspondence to Maxwell
theory for weak fields, but apparently this is a desirable
feature if one intents to obtain charged Lifshitz black holes
with generic anisotropy. An objective that we will be able to  achieve in
the next section.

\section{Solving field equations}

We look for an asymptotically Lifshitz ansatz of the form
\begin{equation}\label{eq:g_ansatz}
ds^2=-\frac{r^{2z}}{l^{2z}}f(r)dt^2+\frac{l^2}{r^2}\frac{dr^2}{f(r)}+r^2d\vec{x}^2,
\end{equation}
where the gravitational potential satisfies $f(\infty)=1$. For
spin-$1$ fields we assume electric ansatze
$B_{\mu}=B_t(r)\delta_{\mu}^t$ and
$P_{\mu\nu}=2\delta_{[\mu}^t\delta_{\nu]}^rD(r)$. We start by
integrating the nonlinear Maxwell equations (\ref{eq:Max})
which give the generalization of the Coulomb law
\begin{equation}\label{eq:solD}
P_{\mu\nu}=2\delta_{[\mu}^t\delta_{\nu]}^r\frac{Q}{l^{D-2}}\left(\frac{l}{r}\right)^{D-z-1}
\quad\longrightarrow\quad
P=-\frac{Q^2}{2r^{2(D-2)}},
\end{equation}
where the integration constant $Q$ is related to the electric
charge, and for spacetime dimension $D$ it has dimension of
$\mathrm{length}^{D-3}$. The corresponding electric field can
be evaluated from the constitutive relations (\ref{eq:consti}),
$E\equiv{F}_{tr}=\mathcal{H}_PD$, after substituting the
Coulomb law above in the structural function (\ref{eq:H}). The Proca field is obtained algebraically from the difference
between the diagonal temporal and radial components of the
mixed version of Einstein equations (\ref{eq:Ein})
\begin{equation}\label{eq:SolProca}
B_t(r)=\sqrt{\frac{(D-2)(z-1)}{\kappa}}\left(\frac{r}{l}\right)^z\frac{f(r)}{lm},
\end{equation}
which is only possible for $z>1$. Substituting now this
expression in the Proca equation (\ref{eq:Proca}) and taking
the gravitational potential as $f=1-h$ we get
\begin{equation}\label{eq:Euler}
r^2h''+(z+D-1)rh'+(D-2)zh=-l^2m^2+(D-2)z,
\end{equation}
which is an Euler differential equation with a constant
inhomogeneity for the function $h$. The function $h$ must
vanish at infinity in order to satisfy the Lifshitz asymptotic;
this implies the vanishing of the constant inhomogeneity which
fixes the value of the Proca field mass to
\begin{equation}\label{eq:m}
m=\frac{\sqrt{(D-2)z}}{l}.
\end{equation}
The general solution of the Euler equation (\ref{eq:Euler})
compatible with the Lifshitz asymptotic is a superposition of two
negative powers of $r$ which are roots of the following
characteristic polynomial, obtained from substituting
$h\propto1/r^\alpha$ in the differential equation,
\begin{equation}
(\alpha-z)(\alpha-D+2)=0.
\end{equation}
Hence, the solution of the gravitational potential for a
generic value $z>1$ of the dynamical critical exponent is
\begin{equation}\label{eq:solf}
f(r)=1-M_1\left(\frac{l}{r}\right)^{D-2}+M_2\left(\frac{l}{r}\right)^z.
\end{equation}
This is true except for $z=D-2$ where there is a double root of
the characteristic polynomial and the single negative power
must be supplemented with a logarithmic behavior; we analyze
this case in the next subsection.

Returning to Einstein equations we choose the parameterization
for the cosmological constant supporting the Lifshitz spacetime
(\ref{eq:Lifshitz}) in the absence of an electromagnetic field
\begin{eqnarray}
\lambda=-\frac{z^2+(D-3)z+(D-2)^2}{2l^2}.
\end{eqnarray}
It is straightforward to show that using all the above
ingredients, e.g.\ substituting the generalized Coulomb law
(\ref{eq:solD}) in the structural function (\ref{eq:H}) and the
result in the energy-momentum tensor, all the remaining
Einstein equations are satisfied if the previous integration
constants are fixed in terms of the charge via the structural
coupling constants as
\begin{equation}\label{eq:Ms2betas}
M_1=\beta_1\frac{|Q|}{l^{D-3}}, \quad
M_2=\beta_2\left(\frac{|Q|}{l^{D-3}}\right)^{\frac{z}{D-2}}.
\end{equation}

Finally, a comment is needed regarding the AdS generalization
$z=1$ of these solutions. It cannot be obtained taking
trivially the limit $z\rightarrow1$. The reason is that as it
can be anticipated from the expression (\ref{eq:SolProca}), the
Proca field is not needed in this case, and this implies that
the equation (\ref{eq:Euler}) no longer determines the
gravitational potential. In other words, there is no constraint
fixing the electrodynamics, and solutions for any of such
theories are possible. In particular, for the case considered
in the paper we recover the terms of the above gravitational
potential in addition to a new term accompanied by a genuine
integration constant related to the standard AdS mass.

\subsection{Nonlinearly charged logarithmic Lifshitz black hole for $z=D-2$
\label{subsec:log}}
As mentioned previously, when the dynamic critical exponent
takes the value $z=D-2$ we find a double multiplicity for the
decay power $D-2$, which forces the incorporation of a
logarithmic behavior for the gravitational potential according
to
\begin{equation}\label{eq:solflog}
f(r)=1-\left(\frac{l}{r}\right)^{D-2}\left[M_1\ln\left(\frac{r}{l}\right)-M_2\right].
\end{equation}
It is interesting and useful to realize the above solution as a
non-trivial limit of the generic case (\ref{eq:solf}). In order
to make clear this limit is possible, we redefine the
integration constants of the generic case by
\begin{equation}\label{eq:Ms2Ms}
\bigl(M_1,M_2\bigr) \mapsto \bigl((z-D+2)M_1,M_2-M_1\bigr),
\end{equation}
which allows to rewrite (\ref{eq:solf}) in the following way
\begin{equation}\label{eq:solfrew}
f(r)=1-\left(\frac{l}{r}\right)^z\left\{\frac{M_1}{z-D+2}
\left[\left(\frac{r}{l}\right)^{z-D+2}-1\right]-M_2\right\}.
\end{equation}
Taking now the limit $z \rightarrow D-2$ and using the
following definition for the logarithmic function
\begin{equation}\label{eq:log}
\ln(x)\equiv\lim_{\alpha\rightarrow0}\frac{x^\alpha-1}{\alpha},
\end{equation}
we obtain the logarithmic Lifshitz potential
(\ref{eq:solflog}). The fact that the gravitational potential
of the critical case $z=D-2$ can be achieved through a limiting
procedure allows to guess its supporting electrodynamics by a
similar reasoning. Using the linear dependence between the
integration and structural coupling constants of the generic
case (\ref{eq:Ms2betas}), the redefinitions (\ref{eq:Ms2Ms})
are equivalent to redefine the structural coupling constants
along the same way
\begin{equation}\label{eq:betas2betas}
\bigl(\beta_1,\beta_2\bigr) \mapsto \bigl((z-D+2)\beta_1,\beta_2-\beta_1\bigr),
\end{equation}
which allows to rewrite the generic structural function
(\ref{eq:H}) as
\begin{eqnarray}
\mathcal{H}(P)&=&
\frac1{2\kappa l^2}\left\{\frac{(D-2)^2}{z-D+2}\left[(-2l^2P)^{\frac{z-D+2}{2(D-2)}}-1\right]
-2z-D+4\right\}\beta_1\sqrt{-2l^2P} \nonumber\\\nonumber\\
&&-\frac{(z-1)}{\kappa z}{\beta_1}^2P
+\frac{(D-2)^2}{2\kappa l^2}\beta_2\left(-2l^2P\right)^{\frac{z}{2(D-2)}}.\label{eq:Hrew}
\end{eqnarray}
Taking again the limit $z \rightarrow D-2$ and using the above
definition for the logarithmic function, we obtain the following
structural function
\begin{eqnarray}
\mathcal{H}(P)&=&\frac1{2\kappa l^2}\left\{
\left[(D-2)\ln\sqrt{-2l^2P}-3D+8\right]\beta_1+(D-2)^2\beta_2\right\}\sqrt{-2l^2P}
\nonumber\\\nonumber\\
&&{}-\frac{(D-3)}{(D-2)\kappa }{\beta_1}^2P,\label{eq:Hlog}
\end{eqnarray}
which is in fact the logarithmic electrodynamics supporting the
solution for $z=D-2$ as it can be checked directly from the
remaining Einstein equations. The fixing of the integration
constants in terms of the charge via the structural coupling
constants can be understood straightforwardly in this process.
However, it is more illustrative to deduce them from the
limiting procedure. Writing the redefined integration constants
(\ref{eq:Ms2Ms}) in terms of the old ones, using the relation
between the last and their structural coupling constants as
corresponds to the generic case (\ref{eq:Ms2betas}), writing
these old coupling constants in terms of the redefined ones
(\ref{eq:betas2betas}), and finally taking the limit $z
\rightarrow D-2$ we obtain the following relations
\begin{equation}\label{eq:Mslog2betas}
M_1=\beta_1\frac{|Q|}{l^{D-3}}, \quad
M_2=\left(\beta_2+\frac{\beta_1}{D-2}\ln\frac{|Q|}{l^{D-3}}\right)\frac{|Q|}{l^{D-3}}.
\end{equation}
These fixings are confirmed by the Einstein equations when
$z=D-2$. Finally, we comment that logarithmic Lifshitz black
holes are rare, the other example we know is produced using
higher curvature gravity \cite{Lif-R2}.

\subsection{Switching on topology}

Here, we show that the previous solutions can be extended to
topological Lifshitz charged black holes that are solutions
with more general constant curvature horizon topologies. This
can be done by changing the $(D-2)$-dimensional flat base
metric $d\vec{x}^2$ in (\ref{eq:g_ansatz}) with the metric
$d\Omega_k^2=\gamma_{ij}(\vec{x})dx^idx^j$ of a space of
constant curvature $k=\pm 1, 0$, i.e.
\begin{equation}\label{eq:g_top_ansatz}
ds^2=-\frac{r^{2z}}{l^{2z}}f(r)dt^2+\frac{l^2}{r^2}\frac{dr^2}{f(r)}+r^2d\Omega_k^2.
\end{equation}
The resulting potentials for the metric and matter fields will
be all the same that the previously studied flat ones ($k=0$),
which at least in the case of the metric is surprising since
the gravitational potential is usually topology dependent. The
only change is that in order to support nontrivial topologies,
$k\neq0$, we need to consider a different nonlinear
electrodynamics by generalizing the structural functions
(\ref{eq:H}) and (\ref{eq:Hlog}) to
\begin{equation}\label{eq:H_k}
\mathcal{H}_k(P)=\mathcal{H}(P)-\frac{(D-2)(D-3)}{2\kappa l^2}
\beta_k\left(-2l^2P\right)^{1/(D-2)}.
\end{equation}
Unfortunately, for nontrivial topologies the charge is no
longer an independent integration constant, as in the flat
case, since it becomes related to the structural coupling
constant of the topological contribution according to
\begin{equation}\label{eq:Q2beta}
\beta_k\left(\frac{|Q|}{l^{D-3}}\right)^{\frac{2}{D-2}}=k.
\end{equation}

\section{Invertible power-law electrodynamics for $\beta_1=0$}

An interesting special case is when the structural coupling
constant $\beta_1$ vanishes, the structural function becomes a
power-law
\begin{equation}\label{eq:Hpower}
\mathcal{H}(P)=
\frac{(D-2)^2}{2\kappa l^2}\beta_2\left(-2l^2P\right)^{\frac{z}{2(D-2)}},
\end{equation}
and the constitutive relations (\ref{eq:consti}) are
invertible. This case can be described by a Lagrangian,
$\mathcal{L}\equiv\frac12P^{\mu\nu}F_{\mu\nu}-\mathcal{H}=2P\mathcal{H}_P-\mathcal{H}$,
which can be written as a standard function of the invariant
$F\equiv\frac{1}{4}F_{\mu\nu}F^{\mu\nu}={\mathcal{H}_P}^2P$
according to
\begin{equation}\label{eq:Lpower}
\mathcal{L}(F)=\frac{(z-D+2)}{l^2z}\bar{\beta}_2
\left(-\frac{2l^2F}{\bar{\beta}_2^2}\right)^{\frac{z}{2(z-D+2)}},
\end{equation}
where we have redefined appropriately the structural coupling
constant as $\bar{\beta}_2=(D-2)z\beta_2/(2\kappa)$. Notice
that for $z=2(D-2)$ and fixing $\bar{\beta}_2=-1$ we recover
Maxwell theory. This kind of power-law Lagrangian has been
exhaustively studied in Refs.~\cite{Fq}. Note that in
Ref.~\cite{Dehghani:2013mba}, the authors consider such
nonlinear electrodynamics and study the implications concerning
the thermodynamics of Lifshitz black holes charged in this way.

\section{Characterizing the black holes}

The gravitational potential corresponding to a generic value of
the dynamical critical exponent $z>1$ is given by
Eq.~(\ref{eq:solf}). Using the relations (\ref{eq:Ms2betas}) for
the integration constants in terms of the charge via the
structural coupling constants, the potential can be written as
\begin{equation}\label{eq:solfrew}
f(r)=1-\beta_1\left(\frac{\hat{l}}{r}\right)^{D-2}
+\beta_2\left(\frac{\hat{l}}{r}\right)^z,
\end{equation}
where the role of the charge is reduced to define a new length
scale $\hat{l}^{D-2}\equiv{l}|Q|$. In fact, using a dynamical
scaling (\ref{eq:anisyymetry}) with
$\lambda^{D-2}\equiv{l}^{D-3}/|Q|$, we can fix $|Q|=l^{D-3}$ (or
equivalently $\hat{l}=l$) in the whole solution. However, we
do not consider this option in order to preserve the anisotropic scaling
symmetry asymptotically. There is no restriction a priori on
the values and signs of the structural coupling constants; this in turn
makes possible that different kinds of black holes are encoded
in that single potential. Each black hole is associated with a
specific electrodynamics. We analyze all these cases in this
section, and also study  the logarithmic case with $z=D-2$.

First, we notice that the potential allows an extremum if
\begin{equation}\label{eq:extremun}
(D-2)\beta_1\left(\frac{\hat{l}}{r}\right)^{D-2}=z\beta_2\left(\frac{\hat{l}}{r}\right)^z,
\end{equation}
which implies that the constants $\beta_1$ and $\beta_2$ must have
the same sign. At such extremum, the second derivative is
evaluated as
\begin{equation}\label{eq:sec_der}
f''=\frac{(D-2)(z-D+2)\beta_1}{\hat{l}^2}\left(\frac{\hat{l}}{r}\right)^D
   =\frac{z(z-D+2)\beta_2}{\hat{l}^2}\left(\frac{\hat{l}}{r}\right)^{z+2},
\end{equation}
i.e.\ the extremum is a minimum if $(z-D+2)\beta_1>0$ [or
equivalently $(z-D+2)\beta_2>0$] and a maximum if
$(z-D+2)\beta_1<0$ [or equivalently $(z-D+2)\beta_2<0$]. In
other words, the two cases allowing a minimum are $\beta_1>0$,
$\beta_2>0$, $z>D-2$ on the one hand and on the other hand
$\beta_1=-\tilde{\beta}_1<0$, $\beta_2=-\tilde{\beta}_2<0$,
$1<z<D-2$; both cases give rise to black holes with inner and
outer horizons. The maximum is achieved also for two cases: the
first one has $\beta_1>0$, $\beta_2>0$, $1<z<D-2$, and the
second one is for the values $\beta_1=-\tilde{\beta}_1<0$,
$\beta_2=-\tilde{\beta}_2<0$, $z>D-2$; they produce black holes
with a single horizon. This is also the case when no extremum
is possible, i.e. $\beta_1>0$ and $\beta_2=-\tilde{\beta}_2<0$,
which occurs for the two cases $z>D-2$ and $1<z<D-2$. Let us
analyze all those cases in details in two separate subsections.
\subsection{Black holes with two horizons}

We start with the minimum cases having $\beta_1>0$, $\beta_2>0$
and $z>D-2$. In this situation, the minimum is achieved at
\begin{equation}\label{eq:r_min}
r_{\mathrm{min}}=\hat{l}\left(\frac{z}{\beta_1}\frac{\beta_2}{D-2}\right)^{1/(z-D+2)},
\end{equation}
and has value
\begin{equation}\label{eq:f_min}
f(r_{\mathrm{min}})=1-(z-D+2)\left[\left(\frac{\beta_1}{z}\right)^z
\left(\frac{D-2}{\beta_2}\right)^{D-2}\right]^{1/(z-D+2)}.
\end{equation}%
\begin{figure}[h]
\begin{center}
\includegraphics[width=0.49\textwidth]{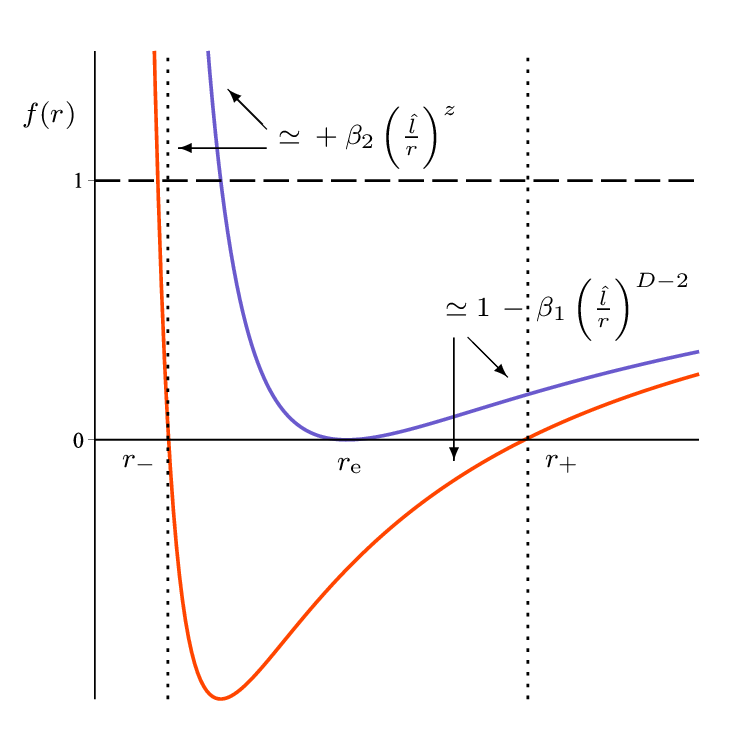}
\includegraphics[width=0.49\textwidth]{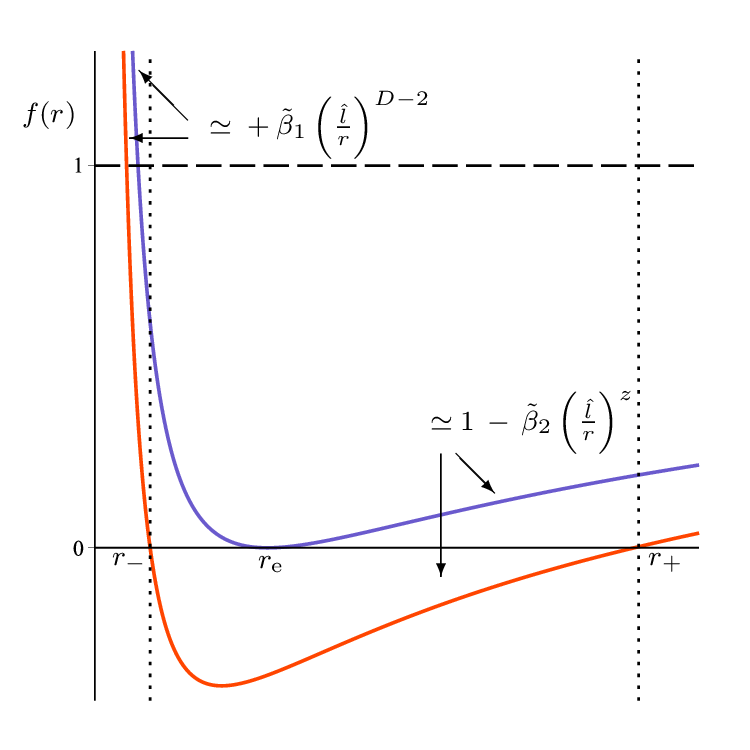}
\end{center}
\caption{These graphs represent the generic behavior of the
gravitational potential $f(r)$ when having a minimum.
In the figure on the left, the red graph represent two-horizons
black holes with dynamic exponent $z=2(D-2)>D-2$
and structure coupling constants set to the values
$\beta_1>0$ and $\beta_2=\frac{1}{2}\,\beta_1^2>0$.
As for the figure on the right it is considered
$\beta_1=-\tilde{\beta}_1<0$,
$\beta_2=-\tilde{\beta}_2=-2\tilde{\beta}_1^{1/2}<0$ and
$z=(D-2)/2<D-2$. The blue graphs represent the extremal black hole,
which is the same in both cases.\label{ML:fig1}}
\end{figure}%
On the other hand, for the analyzed values, the potential goes
asymptotically to one from below as
$1-\beta_1(\hat{l}/r)^{D-2}$ and goes to infinity at $r=0$ as
${}+\beta_2(\hat{l}/r)^z$; we show examples of these behaviors
in the left graph of Figure~\ref{ML:fig1}.
Consequently, we are in the presence of black
holes if $f(r_{\mathrm{min}})\leq0$, i.e. if
\begin{equation}\label{eq:bh_cond}
\left(\frac{\beta_2}{D-2}\right)^{D-2}\leq(z-D+2)^{z-D+2}\left(\frac{\beta_1}{z}\right)^z.
\end{equation}
For the strict inequality, the corresponding black holes have
two horizons $r_{\pm}$ defined by the two zeros of the
potential $f(r_{\pm})=0$. As usual, the event horizon is
defined by the largest of the zeros $r_{+}$. When the
inequality is saturated we obtain an extremal black hole with
zero temperature. The horizon of the extremal black hole is at
\begin{equation}\label{eq:r_e}
r_{\mathrm{e}}=\hat{l}\left(\frac{z-D+2}{z}\beta_1\right)^{1/(D-2)},
\end{equation}
This allows to rewrite the gravitational potential of the
extremal case in a simpler form as
\begin{equation}\label{eq:solf_e}
f(r)=1-\frac{z}{z-D+2}\left(\frac{r_{\mathrm{e}}}{r}\right)^{D-2}
+\frac{D-2}{z-D+2}\left(\frac{r_{\mathrm{e}}}{r}\right)^z.
\end{equation}
Let us now consider the cases allowing a minimum for
$\beta_1=-\tilde{\beta}_1<0$, $\beta_2=-\tilde{\beta}_2<0$ and
$1<z<D-2$, i.e.
\begin{equation}\label{eq:solfrep_tilde}
f(r)=1+\tilde{\beta}_1\left(\frac{\hat{l}}{r}\right)^{D-2}
-\tilde{\beta}_2\left(\frac{\hat{l}}{r}\right)^z.
\end{equation}
These cases are similar to the previous ones, and the main
difference is that now the solutions goes asymptotically to one
from below as $1-\tilde{\beta}_2(\hat{l}/r)^z$ and diverge to
infinity at $r=0$ as ${}+\tilde{\beta}_1(\hat{l}/r)^{D-2}$;
see the right graph of Figure~\ref{ML:fig1}.
The two-horizons black holes exist now for
\begin{equation}\label{eq:bh_cond2}
\left(\frac{\tilde{\beta}_1}{z}\right)^z\leq(D-z-2)^{D-z-2}\left(\frac{\tilde{\beta}_2}{D-2}\right)^{D-2}.
\end{equation}
The extremal case saturating the inequality has exactly the
same extremal horizon and functional form than the previous
one, this last is just more appropriately written as
\begin{equation}\label{eq:solf_e}
f(r)=1+\frac{z}{D-z-2}\left(\frac{r_{\mathrm{e}}}{r}\right)^{D-2}
-\frac{D-2}{D-z-2}\left(\frac{r_{\mathrm{e}}}{r}\right)^z.
\end{equation}
\subsection{Black holes with a single horizon}

The cases with a single horizon contain those two ones where the
extremum is a maximum and other two cases where there is no
extremum at all.

\begin{figure}[h]
\begin{center}
\includegraphics[width=0.49\textwidth]{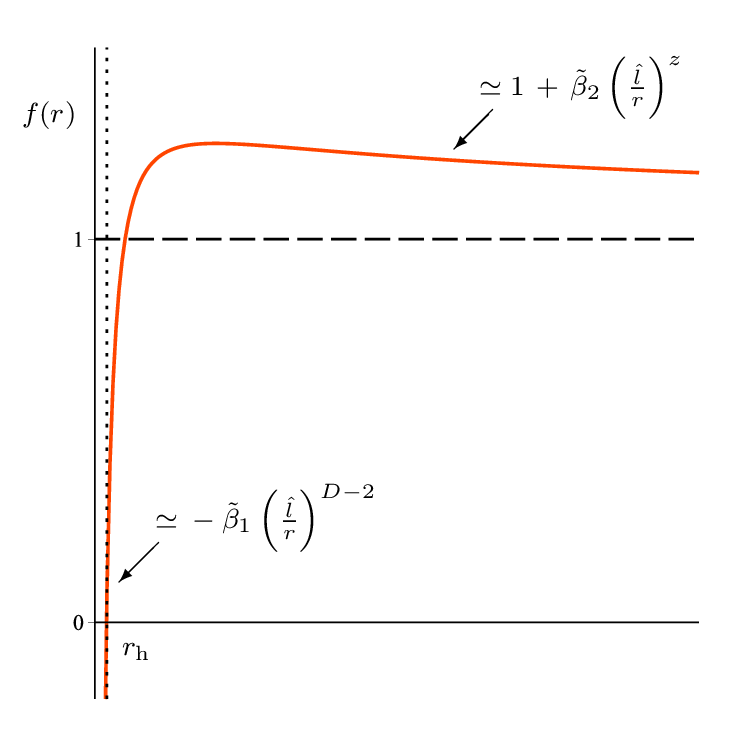}
\includegraphics[width=0.49\textwidth]{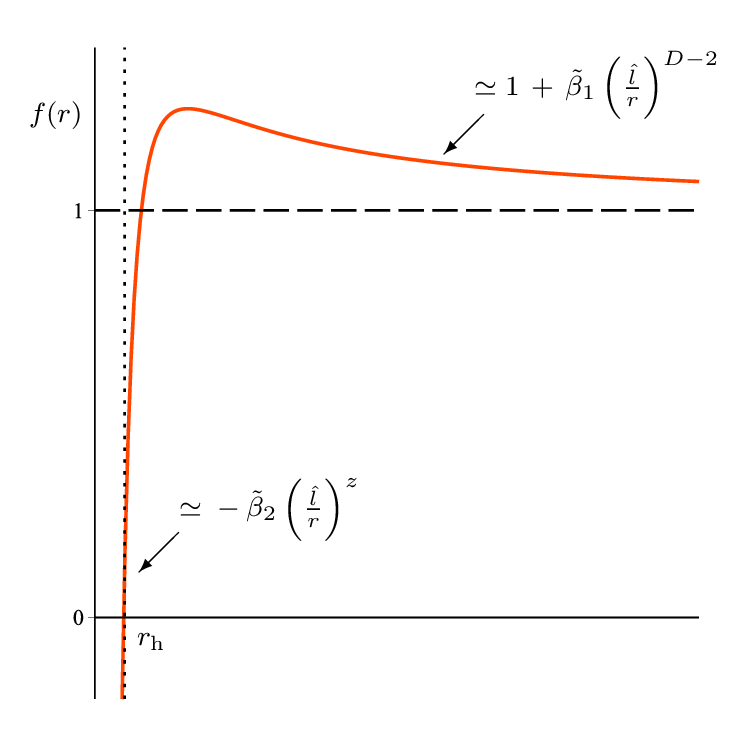}
\end{center}
\caption{The graphs represent the cases with a single horizon
where the extremum is a maximum.
In the graph on the left the dynamic exponent takes the value
$z=(D-2)/2<D-2$ and the considered coupling constants are
$\beta_1>0$ and $\beta_2=\sqrt{\beta_1}>0$.
The graph on the right is for $z=2(D-2)>D-2$, $\beta_1=-\tilde{\beta}_1<0$
and $\beta_2=-\tilde{\beta}_2=-\tilde{\beta}_1^2<0$.
\label{ML:fig2}}
\end{figure}%
Let us start with the maximum options. The first option is
achieved for $\beta_1>0$, $\beta_2>0$ and $1<z<D-2$. Now, the
potential decays asymptotically to one as
$1+\beta_2(\hat{l}/r)^z$ and goes to minus infinity at $r=0$ as
${}-\beta_1(\hat{l}/r)^{D-2}$. In-between,	increasing $r$ the
potential increases, changes the sign at the horizon and keeps
increasing until it achieves a non-vanishing maximum, from
which start the decay to the asymptotic Lifshitz value; see the
left graph of Figure~\ref{ML:fig2}. The second case is for
$\beta_1=-\tilde{\beta}_1<0$, $\beta_2=-\tilde{\beta}_2<0$ [as
in Eq.~(\ref{eq:solfrep_tilde})], but now with $z>D-2$. The
behavior here is similar than the one analyzed just before and
the difference lies in the fact that the asymptotic decay is
now given as $1+\tilde{\beta}_1(\hat{l}/r)^{D-2}$ and the
divergence to minus infinity at $r=0$ goes as
$-\tilde{\beta}_2(\hat{l}/r)^z$; see the right graph of
Figure~\ref{ML:fig2}.

As it was emphasized at the beginning of the section, the cases
with no extremum at all are those with different values of the
constants, the only possibility compatible with the existence
of an horizon is $\beta_1\geq0$ and
$\beta_2=-\tilde{\beta}_2\leq0$, i.e.
\begin{equation}\label{eq:solfrep_mix1}
f(r)=1-\beta_1\left(\frac{\hat{l}}{r}\right)^{D-2}
-\tilde{\beta}_2\left(\frac{\hat{l}}{r}\right)^z.
\end{equation}%
\begin{figure}[h]
\begin{center}
\includegraphics[width=0.49\textwidth]{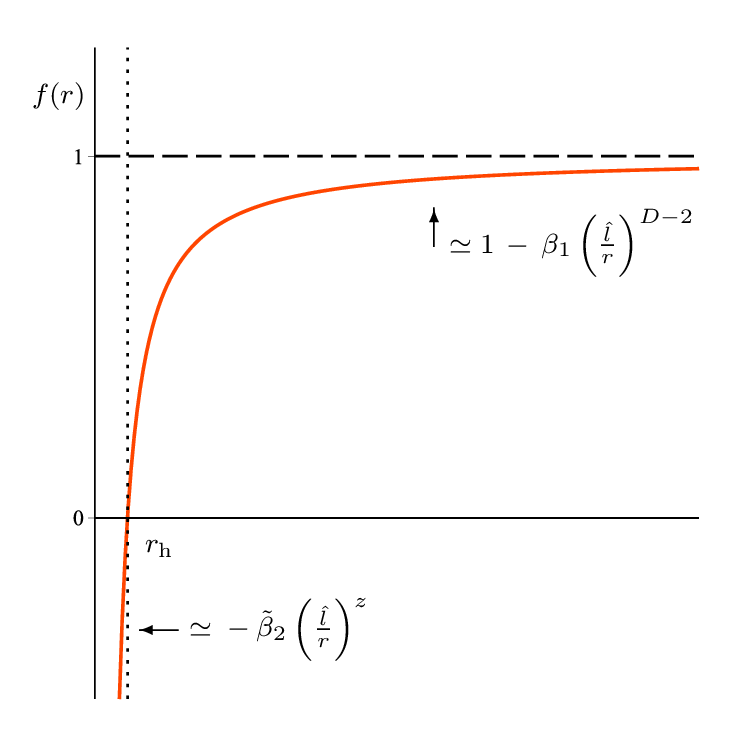}
\includegraphics[width=0.49\textwidth]{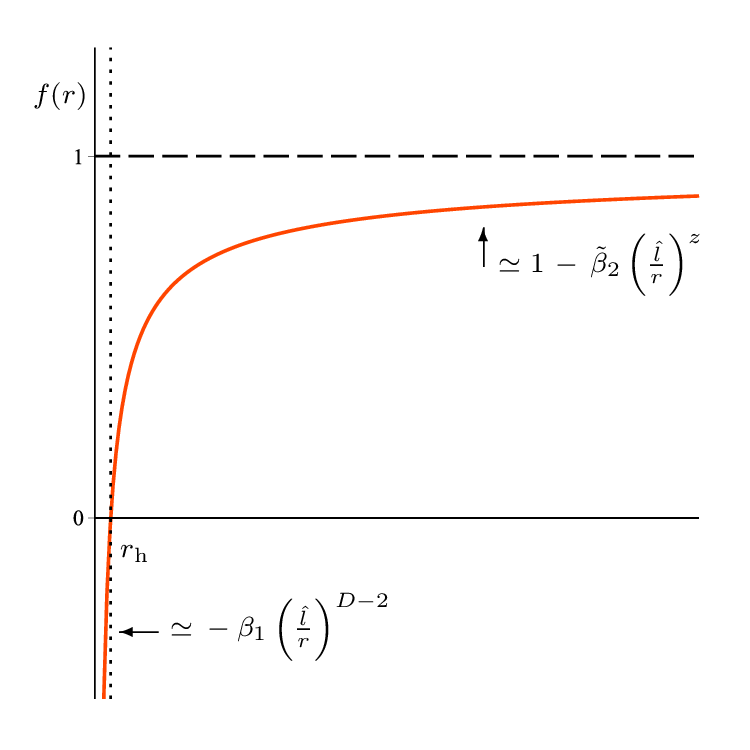}
\end{center}
\caption{These graphics represent black holes with no extremum
for the gravitational potential $f(r)$. In
the left graph $z=2(D-2)>D-2$ with coupling constants $\beta_1>0$ and
$\beta_2=-\tilde{\beta}_2=-\beta_1^2<0$.
The right graph is for $z=(D-2)/2<D-2$, $\beta_1>0$ and
$\beta_2=-\tilde{\beta}_2=-\beta_1^{1/2}<0$.\label{ML:fig3}}
\end{figure}%
The two cases correspond to $z>D-2$ and $1<z<D-2$, since the
behaviors at $r=0$ and at infinity become interchanged, except
when one of the structural couplings constants is zero and
there is no distinction at all. However, in both cases the
solutions are monotonous, diverges to minus infinity at $r=0$,
changes sign at the horizon and keep increasing to achieve the
asymptotically Lifshitz value $f=1$. Both graphs are shown in
Figure \ref{ML:fig3}.

\subsection{Logarithmic black holes}

For $z=D-2$ we obtain the logarithmic potential
(\ref{eq:solflog}), using the fixing of the integration
constants (\ref{eq:Mslog2betas}) it can be written as
\begin{equation}\label{eq:solflogrew}
f(r)=1-\left(\frac{\hat{l}}{r}\right)^{D-2}
\left[\beta_1\ln\left(\frac{r}{\hat{l}}\right)-\beta_2\right].
\end{equation}
Such potential has an extremum at
\begin{equation}\label{eq:extremunlog}
(D-2)\beta_1\ln\left(\frac{r}{\hat{l}}\right)=\beta_1+(D-2)\beta_2,
\end{equation}
where the second derivative is evaluated as
\begin{equation}\label{eq:sec_derlog}
f''=\frac{(D-2)\beta_1}{\hat{l}^2}\left(\frac{\hat{l}}{r}\right)^D,
\end{equation}
which means that the extremum is a minimum for $\beta_1>0$,
giving two-horizons black holes, and a maximum for
$\beta_1=-\tilde{\beta}_1<0$ giving single horizons black
holes.

\begin{figure}[b]
\begin{center}
\includegraphics[width=0.49\textwidth]{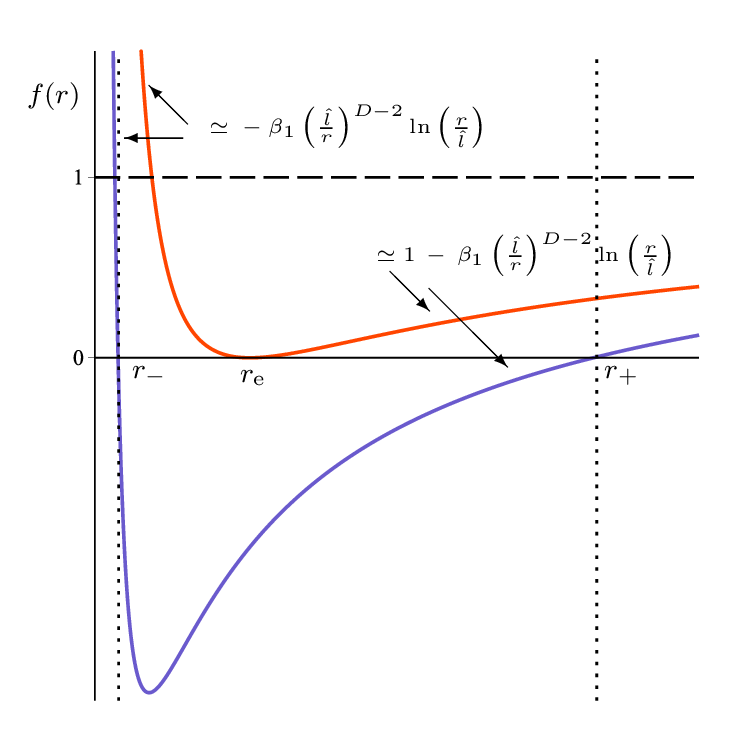}
\includegraphics[width=0.49\textwidth]{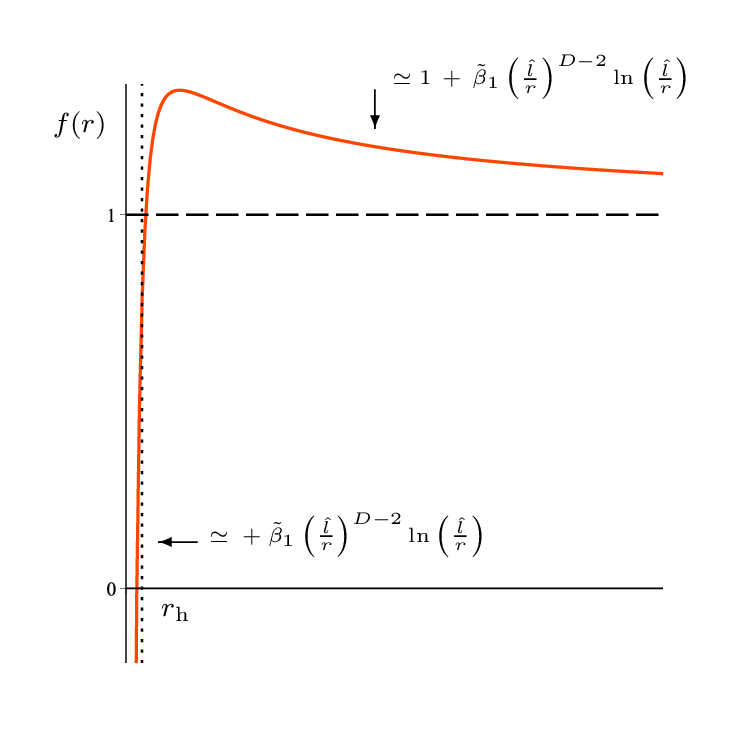}
\end{center}
\caption{The graphics represent logarithmic black holes for $D=5$.
The left graph represents the minimum case, for which $ \beta_1> 0$.
The right graph represents the maximum, where
$\beta_1 = -\tilde{\beta}_1 <0$ and $\beta_2 = - \frac{1}{D-2}\,\beta_1$.
\label{ML:fig4}}
\end{figure}%
For $\beta_1>0$ the minimum achieved at
\begin{equation}\label{eq:r_minlog}
r_{\mathrm{min}}=\hat{l}\exp{\left[\frac{\beta_1+(D-2)\beta_2}{(D-2)\beta_1}\right]},
\end{equation}
has value
\begin{equation}\label{eq:f_minlog}
f(r_{\mathrm{min}})=1-\frac{\beta_1}{D-2}\exp{\left[-1-(D-2)\frac{\beta_2}{\beta_1}\right]}.
\end{equation}
From this minimum the potential grows to the right to achieve
the Lifshitz asymptotic value at infinity as
$1-\beta_1(\hat{l}/r)^{D-2}\ln(r/\hat{l})$ and to the left
increase indefinitely at $r=0$ obeying
${}-\beta_1(\hat{l}/r)^{D-2}\ln(r/\hat{l})$; see the left graph of Figure~\ref{ML:fig4}. The
solution represents black holes if $f(r_{\mathrm{min}})\leq0$,
which occurs for
\begin{equation}\label{eq:bh_condlog}
\beta_2\leq\frac{\beta_1}{D-2}\left[\ln\left(\frac{\beta_1}{D-2}\right)-1\right].
\end{equation}
Notice that using the limit procedure of
Subsec.~\ref{subsec:log} we obtain exactly this inequality from
the one of (\ref{eq:bh_cond}) in the limit
$z\rightarrow(D-2)^+$. When the inequality is saturated we have
an extremal black hole with horizon
\begin{equation}\label{eq:r_elog}
r_{\mathrm{e}}=\hat{l}\left(\frac{\beta_1}{D-2}\right)^{1/(D-2)},
\end{equation}
and the gravitational potential is rewritten as
\begin{equation}\label{eq:solf_elog}
f(r)=1-\left(\frac{r_{\mathrm{e}}}{r}\right)^{D-2}
\left[(D-2)\ln\left(\frac{r}{r_{\mathrm{e}}}\right)+1\right].
\end{equation}

For $\beta_1=-\tilde{\beta}_1<0$ we deal with a potential
having a non-vanishing maximum
\begin{equation}\label{eq:solflogrewmix}
f(r)=1+\left(\frac{\hat{l}}{r}\right)^{D-2}
\left[\tilde{\beta_1}\ln\left(\frac{r}{\hat{l}}\right)+\beta_2\right].
\end{equation}
From this maximum the potential fall to the asymptotic Lifshitz
value to the right according to
$1+\tilde{\beta}_1(\hat{l}/r)^{D-2}\ln(r/\hat{l})$ and to the
left it changes sign at the horizon and continues decreasing
infinitely at $r=0$ as
${}+\tilde{\beta}_1(\hat{l}/r)^{D-2}\ln(r/\hat{l})$; see
the right graph of Figure~\ref{ML:fig4}.

\section{Conclusions}

Here, using the $(\mathcal{H},P)$-formalism, we have proposed a
new nonlinear electrodynamics that together with the Proca
field has allowed us to obtain charged Lifshitz black hole
configurations characterized by the fact that the dynamical
critical exponent can take any value $z>1$. The solutions have
as single integration constant the electric charge, but they
are parameterized by the two arbitrary structural coupling
constant characterizing the family of nonlinear
electrodynamics, in addition to the critical exponent. After
exhaustively studying all the configurations, we conclude that
the family of solutions gives rise to eight different kinds of
black holes; three cases represent two-horizons black holes and
the other five have a single horizon. For a generic exponent
$z\neq{D-2}$ the solutions decay to Lifshitz spacetime by a
negative power, but the value $z=D-2$ turns out to be critical
in the sense that it yields a logarithmic decay supported by a
particular logarithmic electrodynamics. This is the second
example of a logarithmic Lifshitz black hole known in the
literature after the higher-curvature one shown in
Ref.~\cite{Lif-R2}. We provide also a useful limiting procedure
to obtain log configurations and its properties as a nontrivial
limit of the generic ones. For the two asymptotically Lifshitz
behaviors we exhibit examples of extremal black holes, which
were known previously only in the presence of higher-order
theories \cite{Lif-R2}. It is interesting to stress that the
derivation of these solutions has been possible because of the
nonlinear character of the electrodynamics source. It is also
appealing that charged topological Lifshitz black holes can
also be obtained by slightly generalizing the proposed
electrodynamics. Indeed, in this case, the constant curvature
of the horizon is encoded in the nonlinear Lagrangian and does
not appear in the gravitational potential as it usually occurs.
As consequence, the electric charge is no longer an integration
constant if we pursue to incorporate nontrivial horizon
topologies.

It is now accepted that nonlinear electrodynamics enjoy nice
thermodynamical properties since it usually satisfy the zeroth
and first law. It will be nice to study the thermodynamics
issue of these charged solutions in order to give a physical
interpretation of the constants appearing in the solutions. In
Ref.~\cite{Dehghani:2013mba}, the authors explore the possible
thermodynamics behavior of charged Lifshitz black hole
solutions with an electrodynamics given by a power-law of the
Maxwell invariant. However, they do not have
explicit solutions in order to check if their assumptions fit
or not. In Sec.~$4$, we have shown that our electrodynamics
contains these kind of theories as particular case, which
provide an explicit working example.

We are also convinced that nonlinear electrodynamics will still
be useful to explore Lifshitz black hole solutions and its
generalizations in contexts beyond standard gravity, as for
example for the Lovelock gravity or even in higher-order
gravity theories.

\begin{acknowledgments}
This work has been partially supported by grants 1130423,
11090281 and 1121031 from FONDECYT, by grants 175993 and 178346
from CONACyT and by CONICYT, Departamento de Relaciones
Internacionales ``Programa Regional MATHAMSUD 13 MATH-05.''
A.A. is supported by ``Plataforma de Movilidad Estudiantil
Alianza del Pac\'{\i}fico.'' E.A-B is supported by ``Programa
Atracci\'{o}n de Capital Humano Avanzado del Extranjero, MEC'' from
CONICYT. H.G. is supported by IISN-Belgium, and by
``Communaut\'e fran\c{c}aise de Belgique-Actions de Recherche
Concert\'ees.''
\end{acknowledgments}


\end{document}